\documentclass{optica-article}

\journal{opticajournal} 

\articletype{Research Article}

\usepackage{lineno}
\usepackage{placeins}

\begin{document}

\title{Optimization and vectorization of a Mz-type optically-pumped Rubidium magnetometer}

\author{Zhengyu SU,\authormark{1} Yang LI,\authormark{1} Yongbiao YANG,\authormark{1} Yanhua WANG,\authormark{2,3} Jun HE,\authormark{1,3} and Junmin WANG\authormark{1,3*}}

\address{\authormark{1}State Key Laboratory of Quantum Optics Technologies and Devices, and Institute of Opto-Electronics, Shanxi University, Taiyuan, 030006, China\\
\authormark{2}School of Physics and Electronic Engineering, Shanxi University, Taiyuan 030006, China\\
\authormark{3}Collaborative Innovation Center of Extreme Optics, Shanxi University, Taiyuan, 030006, China}

\email{\authormark{*}wwjjmm@sxu.edu.cn} 


\begin{abstract}
Optically pumped magnetometers (OPMs) have demonstrated significant potential in weak magnetic field detection due to their high sensitivity. In this study, we developed an Mz-type optically pumped rubidium magnetometer using a paraffin-coated anti-relaxation vapor cell. The system optimization and performance characterization were conducted inside a magnetic shield. Specifically, the pump light intensity and radio-frequency (RF) magnetic field were jointly optimized by using the linewidth-amplitude ratio as the core metric. Based on the frequency-domain noise spectrum, the sensitivity in open-loop mode was measured to be approximately 30.8~pT/Hz$^{1/2}$. Furthermore, a closed-loop feedback locking technique was applied, reducing the measured noise floor under the tested conditions and improving the sensitivity to $22.9~\mathrm{pT}/\mathrm{Hz}^{1/2}$, with a measured $-3$~dB bandwidth of 123~Hz. The dynamic characteristics were evaluated via magnetic-field step response, showing that the system could track magnetic-field changes stably under closed-loop operation. Finally, by using tri-axial modulation and frequency-domain demodulation, we overcame the scalar measurement limitation of traditional Mz magnetometers. This work realizes vector magnetic field detection and provides a technical basis for applications such as geomagnetic navigation and magnetic anomaly detection.
\end{abstract}

\section{Introduction}

High-sensitivity weak magnetic field detection technology plays a pivotal role in modern science and engineering, particularly in applications such as geomagnetic navigation \cite{ref1,ref2}, geological exploration \cite{ref3}, and biomagnetic signal detection \cite{ref4,ref5} , where it serves an indispensable function. Among various magnetic sensing technologies, while Superconducting Quantum Interference Devices (SQUIDs) offer extremely high sensitivity, they rely on expensive liquid helium cooling equipment. In contrast, Optically Pumped Magnetometers (OPMs) have demonstrated immense application potential \cite{ref6,ref7} due to their advantages of operating without cryogenic cooling, compact size, and high sensitivity, making them widely regarded as the ideal alternative to traditional SQUIDs.

Among the various types of OPMs, the Mz-type magnetometer, based on the light-RF double-resonance principle, has been extensively investigated~\cite{ref8} owing to its compact structure and fast response. Considerable efforts have been devoted to improving its sensitivity. In 2014, Gu \textit{et al.}~\cite{ref9} systematically optimized five key parameters, including the pump light intensity and RF frequency, in a DFB-laser-pumped cesium magnetometer, improving the sensitivity by an order of magnitude. In 2021, Zhang \textit{et al.}~\cite{ref10} introduced a repumping technique that significantly enhanced the amplitude of the $^{85}$Rb magnetic resonance signal. In 2023, Xu \textit{et al.}~\cite{ref11} proposed the linewidth-amplitude ratio (LAR) as an optimization metric for joint parameter tuning, achieving a sensitivity of $1.5~\mathrm{pT}/\mathrm{Hz}^{1/2}$. To address magnetic-noise suppression, in 2024 Zhan \textit{et al.}~\cite{ref12} developed an FPGA-based closed-loop feedback control system, verifying that closed-loop locking can effectively suppress current-noise power spectral density and improve detection performance. More recently, in 2025, Zhu \textit{et al.}~\cite{ref13} and Zhai \textit{et al.}~\cite{ref14} further improved system performance through orthogonal experimental design and vapor-cell size optimization, respectively.

Despite these significant advances, traditional Mz-type magnetometers typically employ buffer gases to suppress spin relaxation. To obtain sufficient atomic number density, the vapor cell often requires heating to high temperatures ($>100$~$^\circ$C). This not only leads to high power consumption and complex thermal insulation but also limits applications involving temperature-sensitive samples. Furthermore, it significantly constrains deployment in scenarios with strict power and volume limitations, such as micro-nano satellite payloads and portable field geomagnetic surveying. To address this, this work develops an $^{87}$Rb atomic magnetometer based on a paraffin-coated anti-relaxation vapor cell. An isotopically enriched $^{87}$Rb vapor cell with deliberately chosen dimensions was adopted to balance spectral purity, interaction volume, and practical compactness. Leveraging the excellent anti-relaxation properties of the coating, the system maintains a long spin coherence time at near room temperature (30--40~$^\circ$C) without the need for buffer gas. However, the long coherence time also renders the atomic system extremely sensitive to variations in pump light intensity and RF field intensity, making the coupling between power broadening and RF saturation effects more pronounced. Drawing on the approach by Xu \textit{et al.}~\cite{ref11}, we propose a joint parameter optimization strategy based on the ``Linewidth-Amplitude Ratio (LAR)'', precisely locating the system's global optimal working point. On this basis, a closed-loop feedback control circuit was constructed to lock the Larmor precession frequency in real time, reduce the measured noise floor under the tested conditions, and improve detection sensitivity. Simultaneously, a magnetic field step response test method was introduced to characterize the system's dynamic tracking capability and locking robustness.

Moreover, traditional Mz-type optical magnetometers are inherently scalar sensors, measuring only the magnitude (modulus) of the total magnetic field, which fails to meet the demand for full vector information in geomagnetic matching navigation and magnetic anomaly detection. Vectorization technology has thus become a research frontier in recent years. As early as 1970, Cohen-Tannoudji \textit{et al.}~\cite{ref15} laid the theoretical foundation for magnetic parametric resonance. Building on this foundation, a series of vectorization schemes have been developed in recent years. In 2021, Xiao \textit{et al.}~\cite{ref16} employed three-axis parametric modulation and lock-in amplification to achieve ultra-high-sensitivity vector measurement under near-zero-field conditions. In 2024, Dawson \textit{et al.}~\cite{ref17} proposed a triaxial vectorization scheme based on harmonic demodulation, while Zou \textit{et al.}~\cite{ref18} reported a rotating-RF-field method for a single-beam optically pumped magnetometer, demonstrating applicability in high-field environments such as the geomagnetic field while retaining a compact architecture favorable for sensor miniaturization. In 2025, Chism \textit{et al.}~\cite{ref19} designed a vectorized rubidium scalar magnetometer for CubeSat applications using dedicated three-axis modulation coils, and Wang \textit{et al.}~\cite{ref20} demonstrated vector magnetic field reconstruction using an alternating fast-rotating field. Nevertheless, many previously reported vectorization schemes still rely on relatively complex structures, specialized modulation configurations, or additional components, which increase implementation difficulty and system cost.

To address this gap, based on the optimized scalar magnetometer, this paper further applies orthogonal low-frequency modulation magnetic fields via tri-axial coils. Using a frequency-domain signal extraction algorithm to resolve the magnetic field vector components, we successfully validated a low-cost, highly integrated vectorization solution based on a single-beam Mz magnetometer within a laboratory magnetic shielding environment. This work provides technical support for applications in complex field scenarios such as geomagnetic surveys.

\section{Theoretical Analysis}
\subsection{Basic Principle of the Mz-type Optically Pumped Magnetometer}
\begin{figure}[htbp]
\centering
\includegraphics[width=\linewidth]{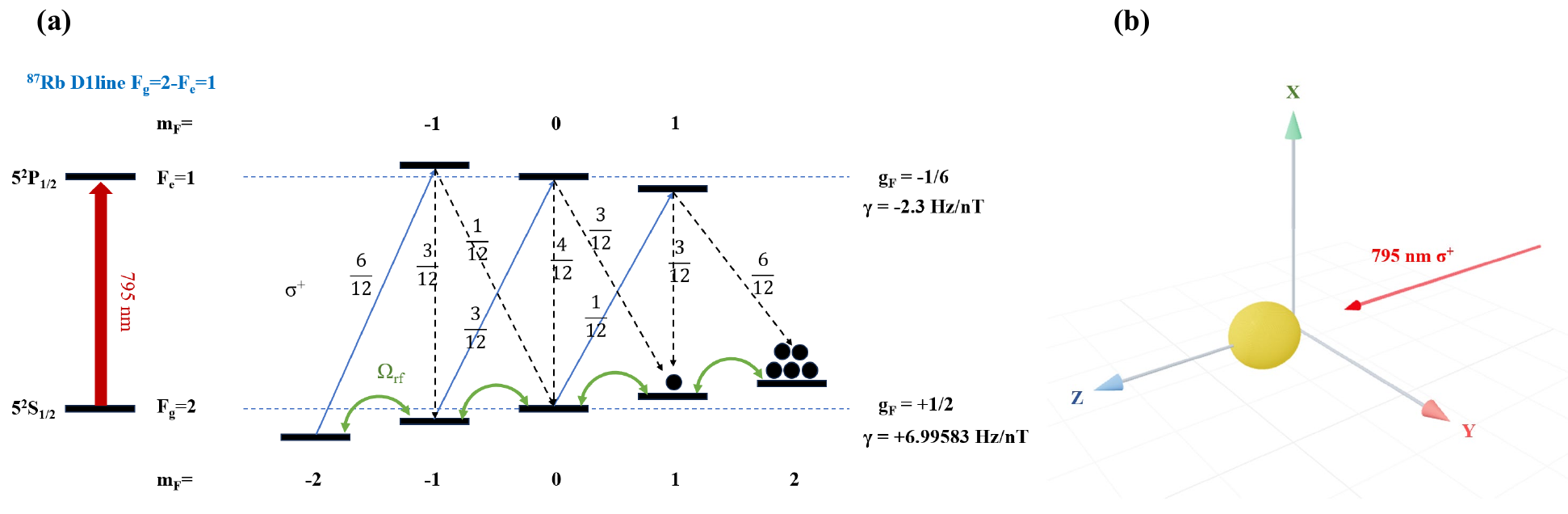} 
\caption{Schematic diagram of the working principle of the $^{87}$Rb Mz atomic magnetometer. \textbf{(a)} Energy level transition diagram of the $^{87}$Rb D1 line. The thick red arrow on the left indicates the D1 transition ($5^2S_{1/2} - 5^2P_{1/2}$) at 795 nm. Solid blue arrows represent the stimulated transitions caused by the absorption of $\sigma^+$ pump light, while dashed black arrows represent the spontaneous emission processes from the excited state back to the ground state. The fractions labeled on the dashed lines indicate the relative transition probabilities (Clebsch-Gordan coefficients) for each path. The curved green arrows ($\Omega_{\text{rf}}$) at the bottom represent the magnetic resonance transitions induced by the radio-frequency field between adjacent Zeeman sublevels. \textbf{(b)} Schematic of the polarization model. The yellow sphere illustrates the atomic ensemble accumulating in the $m_F=+2$ state under the action of optical pumping, resulting in macroscopic spin polarization.}
\label{fig1}
\end{figure}

The operation of the Mz-type optically pumped magnetometer is based on the Zeeman effect and optical pumping technology of alkali metal atoms. As shown in Fig.~\ref{fig1}, the D1 line transition ($5^2S_{1/2} - 5^2P_{1/2}$) of $^{87}$Rb atoms was selected as the working energy level in our experiment. In the absence of an external magnetic field, the ground and excited state energy levels are degenerate, and the population of atoms in each magnetic sublevel follows the Boltzmann distribution, remaining in a state of thermal equilibrium. When a static magnetic field $B_0$ is applied along the $z$-axis, the energy levels split due to the Zeeman effect; specifically, the ground state $F_g=2$ splits into five magnetic sublevels ($m_F = -2, -1, 0, +1, +2$). To break this thermal equilibrium, a beam of circularly polarized light ($\sigma^+$) with a wavelength of 795 nm propagating along the $z$-axis illuminates the vapor cell. According to the law of conservation of angular momentum, when atoms absorb $\sigma^+$ photons and undergo stimulated transitions, the selection rule $\Delta m_F = +1$ must be satisfied. The transition probability is determined by the Clebsch-Gordan (CG) coefficients (as indicated by the values on the dashed lines in Fig.~\ref{fig1}(a)). Atoms in the excited state are unstable and rapidly decay back to the ground state via spontaneous emission (following the $\Delta m_F = 0, \pm 1$ rule). After multiple "optical absorption-spontaneous emission" cycles, since there is no $m_F = +3$ energy level in the excited state $F_e=1$, atoms in the ground state $F_g=2$, $m_F=+2$ cannot absorb $\sigma^+$ photons and are thus trapped in this level. This state is referred to as the "dark state." With the continuous action of the pump light, the majority of atoms eventually accumulate in the $m_F = +2$ state, achieving macroscopic spin polarization of the atoms (as shown in Fig.~\ref{fig1}(b)). At this point, the medium's absorption of the pump light is minimized, macroscopically manifesting as the transmittance of the vapor cell reaching its maximum. Once the atoms are in the polarized state, a radio-frequency (RF) magnetic field $B_{\text{rf}}$ with a frequency of $\Omega_{\text{rf}}$ is applied perpendicular to the optical axis (along the $y$-axis) (denoted as $\Omega_{\text{rf}}$ in Fig.~\ref{fig1}(a)). When the scanning RF field frequency matches the atomic Larmor precession frequency $\omega_L$ (i.e., satisfying $\hbar\omega_{\text{rf}} = \Delta E$), the RF field induces magnetic resonance transitions between adjacent Zeeman sublevels. This causes the atoms originally accumulated in the $m_F = +2$ state to be redistributed to other $m_F$ states, destroying the atomic polarization state (i.e., "depolarization"). The depolarized atoms can re-absorb the pump light, leading to a significant decrease in light intensity transmitted through the cell, forming a distinct absorption dip at the resonance frequency. This constitutes the typical Mz magnetic resonance signal. Based on the resonance condition, by measuring the resonance frequency $\omega_L$ and combining it with the gyromagnetic ratio of the $^{87}$Rb atom ground state ($\gamma = 6.99583$~Hz/nT), the magnitude of the external static magnetic field can be precisely deduced using the formula $B_0 = \omega_L / \gamma$.

\subsection{Signal Model and Sensitivity Optimization Theory}
In the Mz-type optically pumped magnetometer, the dynamic behavior of the macroscopic magnetic moment $\vec{M} = (M_x, M_y, M_z)$ of the alkali metal atomic ensemble, under the combined action of external magnetic fields and relaxation mechanisms, can be described by the Bloch equations:
\begin{equation}
\frac{d\vec{M}}{dt} = \gamma(\vec{M} \times \vec{B}_{\mathrm{tot}}) - \Gamma(\vec{M} - \vec{M}_0) + R_{\mathrm{op}}(M_{\mathrm{max}}\hat{z} - \vec{M})
\end{equation}
where $\gamma$ denotes the gyromagnetic ratio, $\vec{B}_{\mathrm{tot}}$ represents the total magnetic field vector, and $\Gamma$ is the relaxation rate. $R_{\text{op}}$ stands for the optical pumping rate, which is proportional to the pump laser power $P_{\text{pump}}$. It plays a dual role: generating spin polarization while simultaneously introducing an additional power broadening relaxation mechanism.

\FloatBarrier

For the Mz magnetometer, our primary focus is on the magnetic moment component along the optical axis, $M_z$, as it directly determines the vapor cell's absorption characteristics for the probe light (i.e., the transmitted light intensity). The steady-state solution in the rotating coordinate frame reveals that as the RF frequency $\omega$ is scanned across the Larmor frequency $\omega_L$, $M_z$ exhibits a characteristic Lorentzian absorption dip profile:
\begin{equation}
M_z(\delta) = M_0 \left( 1 - \frac{\Omega^2 \frac{\Gamma_2}{\Gamma_1}}{\delta^2 + \Gamma_{\text{eff}}^2} \right).
\label{eq:lorentzian}
\end{equation}
where $\delta = \omega - \omega_L$ represents the frequency detuning between the RF field and the Larmor frequency. $M_0$ denotes the initial macroscopic magnetic moment in the absence of the RF field, which depends on the optical pumping rate $R_{\text{op}}$. $\Omega = \gamma B_{\text{rf}}$ corresponds to the Rabi frequency, characterizing the intensity of the RF magnetic field. $\Gamma_1, \Gamma_2$ are the longitudinal and transverse relaxation rates, respectively; both include the intrinsic relaxation rate $\Gamma_{\text{rel}}$ and the broadening term induced by the optical pumping rate $R_{\text{op}}$.

The sensitivity of the magnetometer, $\delta B$, is determined by the ratio of the full width at half maximum (FWHM) of the magnetic resonance linewidth, $\Delta f$, to the signal-to-noise ratio (SNR):
\begin{equation}
\delta B \propto \frac{\Delta f}{S_{\text{amp}}} \propto \frac{\Gamma_{\text{eff}}}{S_{\text{amp}}},
\label{eq:sensitivity}
\end{equation}
Based on the analytical expression of $M_z(\delta)$, we can derive the mathematical expressions for the signal amplitude $S_{\text{amp}}$ and the effective linewidth $\Gamma_{\text{eff}}$, and analyze their dependence on experimental parameters. The signal amplitude $S_{\text{amp}}$ corresponds to the depth of the absorption dip at the resonance point ($\delta=0$):
\begin{equation}
S_{\text{amp}} = M_z(\infty) - M_z(0) = M_0 \frac{\Omega^2}{\Omega^2 + \Gamma_1\Gamma_2}.
\label{eq:amplitude}
\end{equation}
The optical pumping rate $R_{\text{op}}$ is determined by the pump light intensity. As the light intensity increases, the initial magnetization $M_0$ grows (indicating a larger number of polarized atoms), which consequently enhances the signal amplitude $S_{\text{amp}}$. The RF magnetic field intensity determines the Rabi frequency $\Omega$. As $B_{\text{rf}}$ increases, the $\Omega^2$ term in the numerator becomes dominant, leading to an increase in $S_{\text{amp}}$. However, when the condition $\Omega^2 \gg \Gamma_1\Gamma_2$ is met, the signal tends to saturate. 

The effective linewidth corresponds to the half-width at half-maximum (HWHM) of the Lorentzian curve:
\begin{equation}
\Gamma_{\text{eff}} = \sqrt{\Gamma_2^2 + \frac{\Gamma_2}{\Gamma_1}\Omega^2}.
\label{eq:linewidth}
\end{equation}
Considering the linear dependence of the relaxation rate on the optical pumping rate, $\Gamma_2 \approx \Gamma_{\text{rel}} + \alpha P_{\text{pump}}$, substituting this into the above equation reveals that $\Gamma_{\text{eff}}$ exhibits a linear growth trend with the pump power $P_{\text{pump}}$. Excessive pump power will directly result in spectral line broadening. Meanwhile, $\Gamma_{\text{eff}}$ shows a nonlinear increase with the RF intensity $\Omega$; when $\Omega$ becomes excessive, the linewidth broadens significantly. Synthesizing the two equations above, it is evident that the sensitivity $\delta B$ does not vary monotonically with these parameters. Increasing the pump power enhances $S_{\text{amp}}$, but simultaneously increases $\Gamma_{\text{eff}}$ via the $\Gamma_2$ term. Similarly, increasing the RF field strength enhances $S_{\text{amp}}$, yet it induces saturation and broadening through the $\Omega$ term. Therefore, theoretically, there exists an optimal combination of $P_{\text{pump}}$ and $B_{\text{rf}}$ that minimizes the ratio $\Gamma_{\text{eff}} / S_{\text{amp}}$ (corresponding to the highest sensitivity). This derivation explicitly validates the necessity and scientific rationality of employing a multi-parameter joint optimization strategy.
\subsection{Principle of Vector Measurement via Tri-axial Modulation}

Traditional Mz-type optically pumped magnetometers are inherently scalar sensors, as their output signal responds only to the magnitude (modulus) of the total magnetic field. To achieve vector measurement, this study adopts a vectorization scheme based on auxiliary magnetic field modulation. This method establishes a mapping relationship between the scalar magnitude and the vector components by applying weak low-frequency modulation magnetic fields along three orthogonal axes.

Assume the static magnetic field to be measured is $\vec{B}_0$, with components ($B_x, B_y, B_z$) in the Cartesian coordinate system, and its scalar magnitude is denoted as $|\vec{B}_0|$. To extract directional information, we apply sinusoidal modulation magnetic fields with frequencies $\omega_x, \omega_y, \omega_z$ and a uniform amplitude $\beta$ along the $x, y$, and $z$ orthogonal axes, respectively. Consequently, the total instantaneous magnetic field vector $\vec{B}_{\text{tot}}(t)$ sensed by the atomic vapor cell is the vector sum of the static magnetic field and the modulation magnetic fields:
\begin{equation}
\vec{B}_{\text{tot}}(t) = \vec{B}_0 + \vec{B}_{\text{mod}}(t) = \vec{B}_0 + \sum_{i=x,y,z} \beta \cos(\omega_i t)\vec{e}_i,
\label{eq:vector_sum}
\end{equation}
where $\vec{e}_i$ denotes the unit vector along each axis ($i=x, y, z$), and the modulation depth satisfies the perturbation condition $\beta \ll |\vec{B}_0|$.

The detection signal of the Mz magnetometer corresponds to the scalar magnitude $|\vec{B}_{\text{tot}}(t)|$ of the total magnetic field. According to the definition of the vector magnitude:
\begin{equation}
|\vec{B}_{\text{tot}}(t)| = \sqrt{\vec{B}_{\text{tot}}(t) \cdot \vec{B}_{\text{tot}}(t)} = \sqrt{|\vec{B}_0|^2 + 2\vec{B}_0 \cdot \vec{B}_{\text{mod}}(t) + |\vec{B}_{\text{mod}}(t)|^2}.
\label{eq:magnitude_def}
\end{equation}
By substituting each component into the expression and expanding it, given the condition $\beta \ll |\vec{B}_0|$, the second-order infinitesimal of the modulation field, $|\vec{B}_{\text{mod}}(t)|^2$, can be neglected. Utilizing the first-order Taylor expansion approximation, $\sqrt{A^2+x} \approx A + x/(2A)$, the magnitude of the total magnetic field can be approximately expressed as:
\begin{equation}
|\vec{B}_{\text{tot}}(t)| \approx |\vec{B}_0| + \frac{\vec{B}_0 \cdot \vec{B}_{\text{mod}}(t)}{|\vec{B}_0|}.
\label{eq:taylor}
\end{equation}
Further expand the dot product terms into component form:
\begin{equation}
|\vec{B}_{\text{tot}}(t)| \approx |\vec{B}_0| + \sum_{i=x,y,z} \left( \frac{B_i}{|\vec{B}_0|} \beta \right) \cos(\omega_i t).
\label{eq:component_expand}
\end{equation}
The above equation clearly demonstrates that the output signal of the scalar magnetometer comprises a DC component $|\vec{B}_0|$ (i.e., the magnitude of the static magnetic field) as well as three AC modulation components with frequencies of $\omega_x, \omega_y, \omega_z$, respectively.

According to the principles of spectral analysis, the spectral amplitude $h_i$ corresponding to the modulation frequency $\omega_i$ can be expressed as:
\begin{equation}
h_i = \frac{B_i}{|\vec{B}_0|} \beta,
\label{eq:spectral_amp}
\end{equation}
where $B_i$ represents the scalar component of the static magnetic field along the $i$-axis. Consequently, the inversion formula for each vector component can be derived as:
\begin{equation}
B_i = \frac{h_i \cdot |\vec{B}_0|}{\beta} \quad (i = x, y, z).
\label{eq:inversion}
\end{equation}
This equation indicates that the spectral amplitude $h_i$ is directly proportional to the projected magnetic field component $B_i$ along the corresponding axis, while being inversely proportional to the total magnetic field magnitude $|\vec{B}_0|$. Experimentally, by extracting $h_x, h_y$, and $h_z$ via Fast Fourier Transform (FFT), and combining them with the total magnitude $|\vec{B}_0|$ measured by the magnetometer and the known modulation depth $\beta$, the three-dimensional vector $\vec{B}_0$ of the static magnetic field can be uniquely determined.
\section{Experimental Setup}

\begin{figure}[htbp]
\centering
\includegraphics[width=\linewidth]{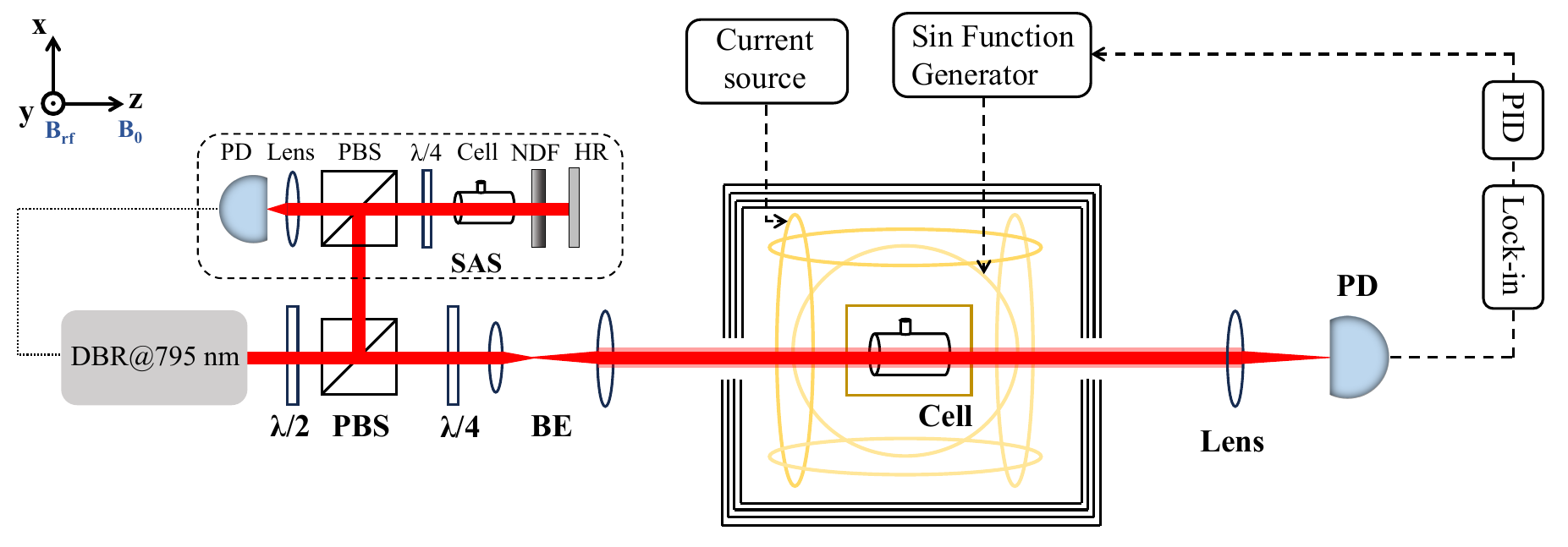} 
\caption{Schematic diagram of the experimental setup for the Mz-type Rb atomic magnetometer. Cell: Paraffin-coated $^{87}$Rb vapor cell; PD: Photodetector; NDF: Neutral density filter; HR: High-reflectivity mirror; PID: Proportional-integral-derivative controller; Lock-in: Digital lock-in amplifier. The vapor cell is placed inside a four-layer permalloy magnetic shield, where tri-axial Helmholtz coils provide the static magnetic field ($B_0$), the radio-frequency field ($B_{\text{rf}}$), and the modulation magnetic fields required for vector measurement.}
\label{fig2}
\end{figure}

Figure~\ref{fig2} illustrates the schematic diagram of the experimental setup for the Mz-type optically pumped rubidium atomic magnetometer. The core sensing unit is a cylindrical high-borosilicate glass vapor cell with a diameter of 25~mm and a length of 75~mm. The cell is filled with isotopically enriched $^{87}$Rb rather than natural-abundance rubidium, in order to suppress the resonance overlap and additional optical absorption arising from $^{85}$Rb and thereby obtain a cleaner working resonance for optical pumping and frequency locking~\cite{ref21}. The cell dimensions were selected as an engineering compromise among interaction volume, wall-relaxation suppression, and compact implementation, rather than as a universal optimum~\cite{ref7,ref14}. In particular, the relatively long cell length helps provide a sufficiently long interaction path and enough participating atoms under near-room-temperature operation. The inner wall of the cell is coated with a paraffin anti-relaxation coating. The paraffin coating effectively suppresses the destruction of spin polarization during collisions between alkali metal atoms and the glass wall, thereby extending the spin coherence time. To maintain an optimal operating temperature, two flexible non-magnetic electric heating films were symmetrically attached to the side walls of the cell to ensure heating uniformity. The heating films were driven by a high-frequency alternating current at 477~kHz to eliminate electromagnetic interference generated by the heating current. Furthermore, the cell temperature was stably maintained at 40~$^\circ$C via a closed-loop temperature control system.

The light source system employed a Distributed Bragg Reflector (DBR) laser with a center wavelength of 795~nm. Using a Saturated Absorption Spectroscopy (SAS) frequency stabilization setup, the laser frequency was precisely locked to the $F_g=2 \rightarrow F_e=1$ hyperfine transition peak of the $^{87}$Rb D1 line. After the laser output, the light intensity was adjusted via a half-wave plate ($\lambda/2$) and a Polarization Beam Splitter (PBS), and the beam was subsequently converted into circularly polarized light by a quarter-wave plate ($\lambda/4$). Considering the 25~mm diameter of the vapor cell, the pump beam was collimated and expanded by a Keplerian telescope system composed of lenses to a spot diameter of approximately 20~mm, so as to improve the overlap between the pump light and the active vapor volume, thereby increasing the effective number of atoms participating in the optical pumping process while retaining a margin to avoid beam clipping at the cell edge. The expanded circularly polarized pump light was incident upon the vapor cell along the $z$-axis to achieve optical pumping spin polarization of the atoms.

To isolate interferences from the geomagnetic field and environmental magnetic noise, the vapor cell was placed at the center of a four-layer permalloy magnetic shield. A set of tri-axial double-wound Helmholtz coils was installed inside the shield to generate the required magnetic field environment. The static magnetic field ($B_0$) was generated by the $z$-axis coil driven by a B2961A precision current source, with its direction aligned along the light wave vector ($z$-axis). The radio-frequency field ($B_{\text{rf}}$) was produced by the $y$-axis coil driven by the signal generation module of a digital lock-in amplifier (MFLI, Zurich Instruments, 5~MHz). Its direction was perpendicular to the optical axis ($y$-axis), serving to drive magnetic resonance transitions. In the vector measurement mode, these tri-axial coils were also utilized to apply orthogonal low-frequency modulation magnetic fields.

The light beam transmitted through the cell was collected by a photodetector. The magnetic resonance signal output from the detector was fed into the digital lock-in amplifier for demodulation. The demodulated error signal (zero-crossing signal) was processed by an internal PID controller to feedback-control the frequency of the RF signal source. Through this closed-loop feedback mechanism, the system locked the RF field frequency to the Larmor precession frequency corresponding to the static magnetic field in real-time, thereby realizing real-time tracking and measurement of the magnetic field.
\section{Experimental Results and Analysis}

\subsection{Parameter Joint Optimization}

To achieve the optimal detection sensitivity of the magnetometer, it is essential to determine the optimal operating points for the pump laser power and the RF magnetic field intensity. Due to the long spin coherence time of the paraffin-coated vapor cell used in the experiment, the system performance is extremely sensitive to variations in light intensity and RF field. Traditional single-variable scanning methods struggle to address the coupling relationship between these two parameters. Therefore, this study adopts a multi-parameter joint optimization strategy. Under a constant cell temperature of 40~$^\circ$C, the linewidth-amplitude ratio (LAR) was utilized as the key figure of merit for evaluating the signal-to-noise ratio (SNR) to locate the system's global optimal working point.

\begin{figure}[htbp]
\centering
\includegraphics[width=0.95\linewidth]{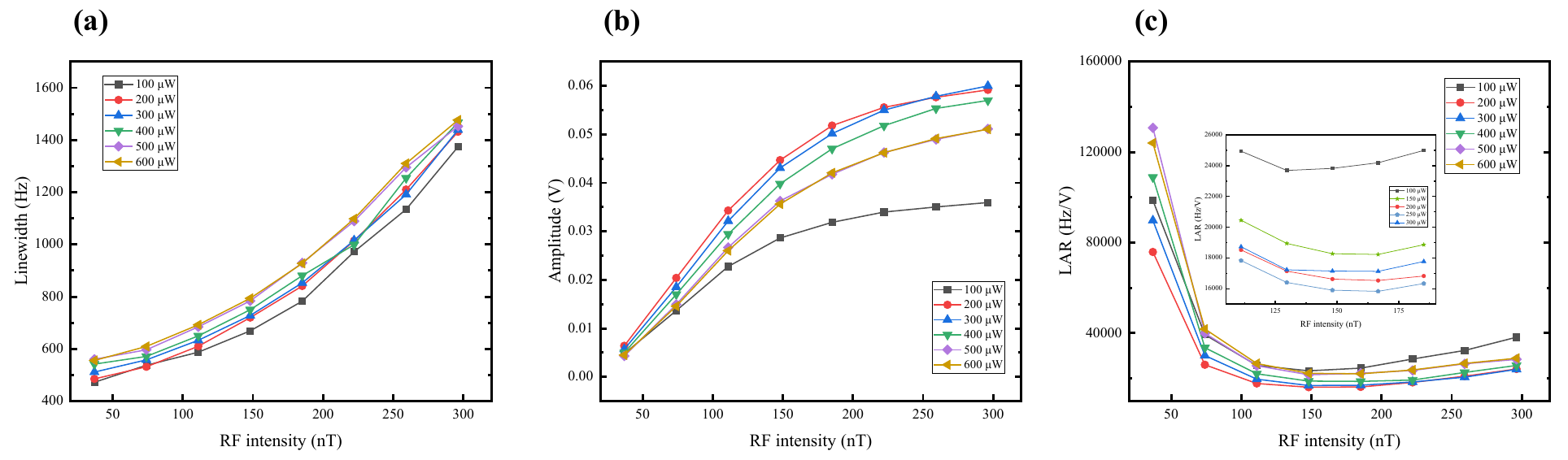} 
\caption{Impact of radio-frequency (RF) field intensity on the characteristics of the magnetic resonance signal. Different colored curves correspond to different pump laser powers ranging from 100~$\mu$W to 600~$\mu$W. \textbf{(a)} Dependence of magnetic resonance linewidth on RF magnetic field. \textbf{(b)} Dependence of signal amplitude on RF magnetic field. \textbf{(c)} Dependence of the Linewidth-Amplitude Ratio (LAR) on RF magnetic field. The inset displays the fine-scan results around the minimum (125--175~nT), used to precisely locate the optimal RF magnetic field ($\sim$165~nT).}
\label{fig3}
\end{figure}

The experiment first characterized the influence of RF magnetic field on the magnetic resonance signal. We set the pump laser power to distinct fixed values within the range of 100~$\mu$W to 600~$\mu$W and scanned the RF magnetic field (ranging from 30~nT to 300~nT). The results are presented in Fig.~\ref{fig3}. Fig.~\ref{fig3}(a) indicates that as the RF magnetic field increases, the magnetic resonance linewidth exhibits a linear growth trend. This is primarily attributed to RF broadening, where a strong RF magnetic field shortens the atomic transverse relaxation time. Fig.~\ref{fig3}(b) shows that the signal amplitude rises rapidly with increasing RF strength but tends to plateau at high field intensities ($>200$~nT) due to the gradual saturation of atomic transitions. Synthesizing these variations, Fig.~\ref{fig3}(c) illustrates the variation curves of LAR with respect to the RF field. It can be observed that the LAR curves under all light intensities exhibit a trend of initially decreasing and subsequently increasing. This reveals the existence of an optimal RF magnetic field interval (150--180~nT) where the LAR reaches a minimum value. To precisely locate this extremum point, a fine scan with reduced step size was performed within this interval (as shown in the inset of Fig.~\ref{fig3}(c)), finally determining the optimal RF magnetic field to be approximately 165~nT.

\begin{figure}[htbp]
\centering
\includegraphics[width=0.95\linewidth]{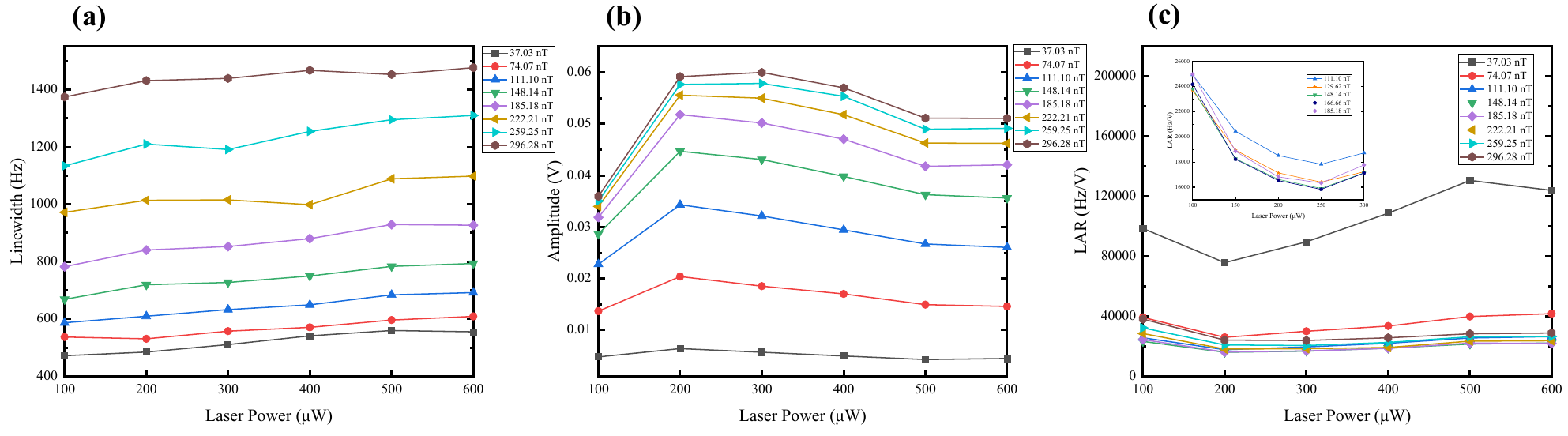} 
\caption{Impact of pump laser power on the characteristics of the magnetic resonance signal. Different colored curves correspond to different radio-frequency (RF) field intensities ranging from 37.03~nT to 296.28~nT. \textbf{(a)} Dependence of magnetic resonance linewidth on pump laser power. \textbf{(b)} Dependence of signal amplitude on pump laser power. \textbf{(c)} Dependence of the Linewidth-Amplitude Ratio (LAR) on pump laser power. The inset displays the fine-scan results under the optimal RF field, identifying the optimal pump laser power at approximately 250~$\mu$W.}
\label{fig4}
\end{figure}

Subsequently, the influence of pump laser power (ranging from 100 to 600~$\mu$W) on system performance was systematically characterized under different RF magnetic field (30--300~nT), as illustrated in Fig.~\ref{fig4}. Fig.~\ref{fig4}(a) indicates that with the increase of pump laser power, the magnetic resonance linewidth exhibits a slight broadening trend. Although the magnitude of this variation is smaller compared to RF broadening, it still reflects the power broadening effect induced by the increased optical pumping rate. Fig.~\ref{fig4}(b) reveals that the signal amplitude displays a distinct characteristic of initially increasing and subsequently decreasing with respect to light intensity: at low light intensities, the amplitude rises rapidly with increasing power, reaching a peak at approximately 200~$\mu$W; subsequently, as the light intensity increases further, the signal amplitude exhibits a slow decay due to optical pumping saturation and inhomogeneous light broadening. The LAR curves in Fig.~\ref{fig4}(c) demonstrate that excessively low light intensity results in weak signals, whereas excessively high intensity leads to reduced amplitude coupled with linewidth broadening. Observation of the LAR curves suggests that the optimal optical power range lies between 100~$\mu$W and 300~$\mu$W. Through the fine scan analysis shown in the inset, the optimal pump laser power under the optimal RF field was determined to be approximately 250~$\mu$W.

\begin{figure}[htbp]
\centering
\includegraphics[width=0.6\linewidth]{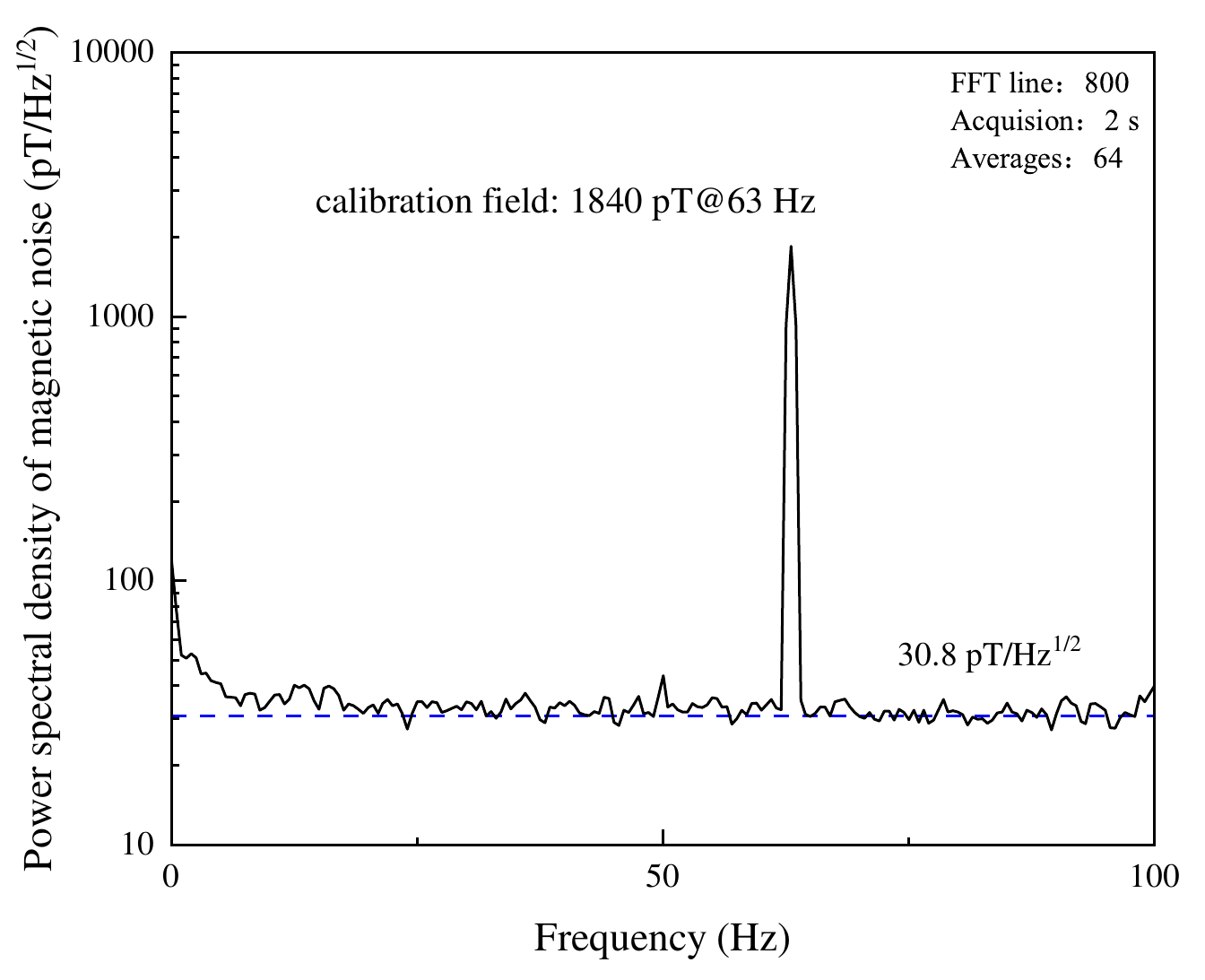} 
\caption{Power spectral density of magnetic noise of the magnetometer in open-loop mode. A sinusoidal calibration magnetic field with a frequency of 63~Hz and an amplitude of 1.84~nT was applied. The sharp peak represents the response to the calibration signal, while the blue dashed line indicates the average noise floor of the system. Based on the signal-to-noise ratio calculation from the calibration signal, the open-loop sensitivity at this operating point is 30.8~pT/Hz$^{1/2}$.}
\label{fig5}
\end{figure}

In summary, through the LAR-based joint optimization strategy, the optimal operating parameters of the magnetometer were determined to be: RF magnetic field intensity $B_{\text{rf}} = 165$~nT and pump laser power $P_{\text{pump}} = 250$~$\mu$W. The corresponding static magnetic field was set to $B_0 = 1216$~nT. At this optimal working point, the sensitivity of the magnetometer was calibrated by applying a weak alternating calibration magnetic field (1.84~nT at 63~Hz). The measured Power Spectral Density (PSD) is presented in Fig.~\ref{fig5}, yielding a sensitivity of 30.8~pT/Hz$^{1/2}$.

\FloatBarrier

\subsection{Closed-Loop Locking and Dynamic Response}

\begin{figure}[htbp]
\centering
\includegraphics[width=0.95\linewidth]{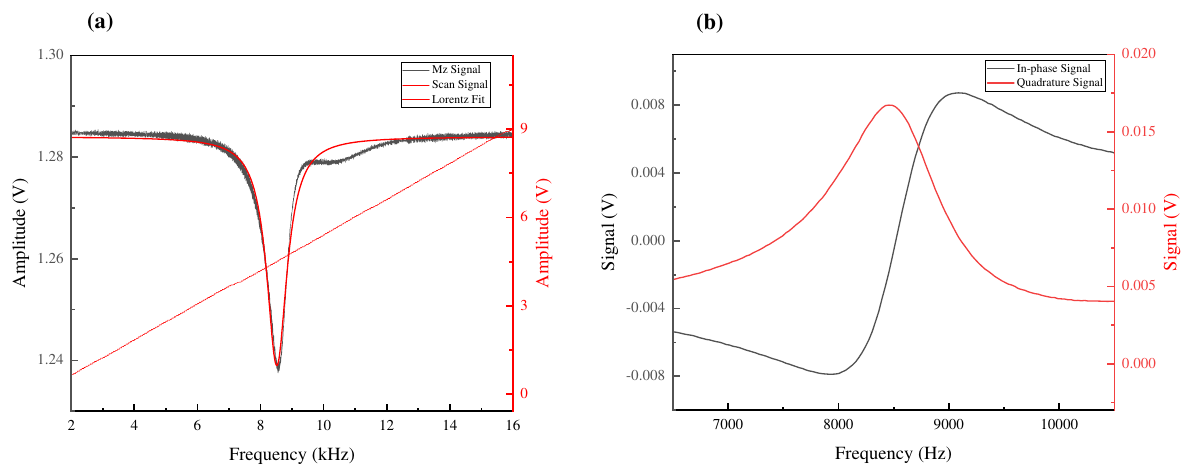} 
\caption{Typical Mz magnetic resonance signal and its demodulated error signal. \textbf{(a)} Magnetic resonance absorption signal observed by scanning the radio frequency (RF) in open-loop mode. When the RF frequency matches the Larmor frequency, the transmitted light intensity decreases significantly, exhibiting a standard Lorentzian lineshape. \textbf{(b)} Error signal (frequency discrimination curve) obtained after demodulation by the lock-in amplifier. The signal displays a distinct linear zero-crossing characteristic (S-curve) near the resonance center. This zero-crossing point serves as the locking reference for the PID controller to achieve closed-loop operation of the magnetometer.}
\label{fig6}
\end{figure}

To achieve real-time continuous measurement of the external magnetic field, the system was switched from the open-loop scanning mode to the closed-loop locking mode. Fig.~\ref{fig6}(a) displays the Mz magnetic resonance signal obtained by scanning the RF frequency under open-loop conditions, which exhibits a typical Lorentzian lineshape. Fig.~\ref{fig6}(b) presents the error signal obtained after demodulation by the lock-in amplifier. In the vicinity of the resonance frequency, the error signal exhibits a distinct linear zero-crossing characteristic, serving as an excellent frequency discrimination curve. We utilized a PID controller to lock the system to the zero-crossing point of this error signal. When variations in the external magnetic field induce a drift in the Larmor frequency, the PID controller adjusts the RF frequency in real-time to maintain the locked state, thereby achieving precise tracking of the magnetic field changes.

\begin{figure}[htbp]
\centering
\includegraphics[width=0.55\linewidth]{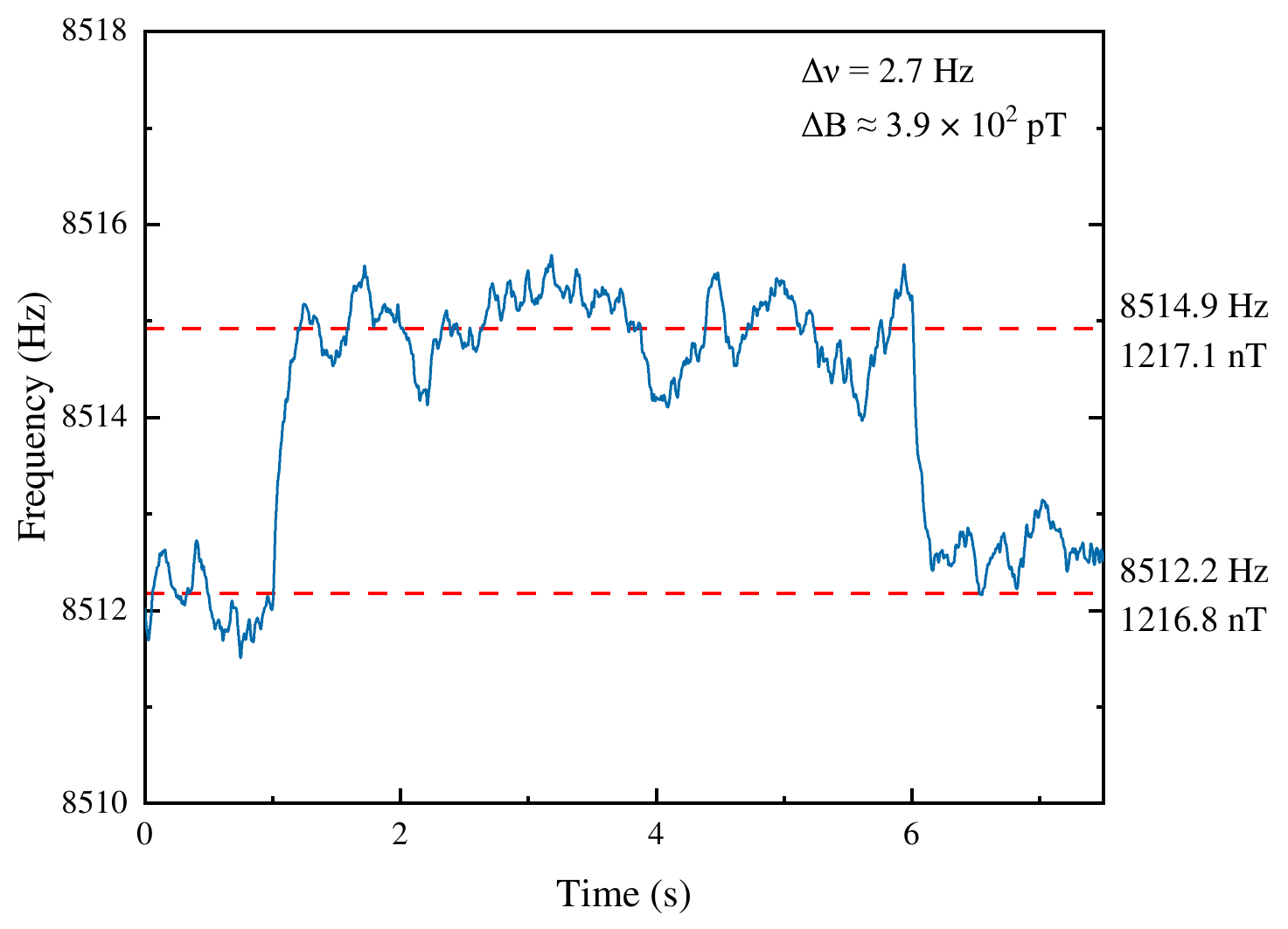} 
\caption{Magnetic field step response test under closed-loop locking conditions. A step magnetic field with an amplitude of approximately 430~pT was applied (turned on at $t = 1$~s and turned off after a duration of 5~s). The results show that the system can accurately track the magnetic field mutation, demonstrating stable closed-loop tracking under the tested step perturbation.}
\label{fig7}
\end{figure}

Upon successfully implementing the closed-loop locking, we first evaluated the system's tracking capability for dynamic magnetic field variations. A square-wave current with a duration of 5~s was applied to the $z$-axis coil to generate a step magnetic field with an amplitude of approximately 430~pT. This step signal was initiated at $t=1$~s and automatically terminated at $t=6$~s. The transient response of the system was characterized by monitoring the real-time variations in the magnetometer's output frequency. The measurement results are depicted in Fig.~\ref{fig7}, where the blue solid line represents the real-time frequency signal output by the magnetometer. It can be observed that upon the application of the step magnetic field, the system's output frequency exhibited a step change of approximately 3.02~Hz. Considering the gyromagnetic ratio of $^{87}$Rb atoms, this frequency shift corresponds to a magnetic field variation of approximately 430~pT. Furthermore, the system responded rapidly to the abrupt magnetic field change. During the step-holding phase, the frequency remained stable without significant overshoot or loss-of-lock phenomena. These results strongly validate that the closed-loop control circuit of the proposed magnetometer possesses excellent dynamic characteristics and robustness.

\begin{figure}[htbp]
\centering
\includegraphics[width=0.5\linewidth]{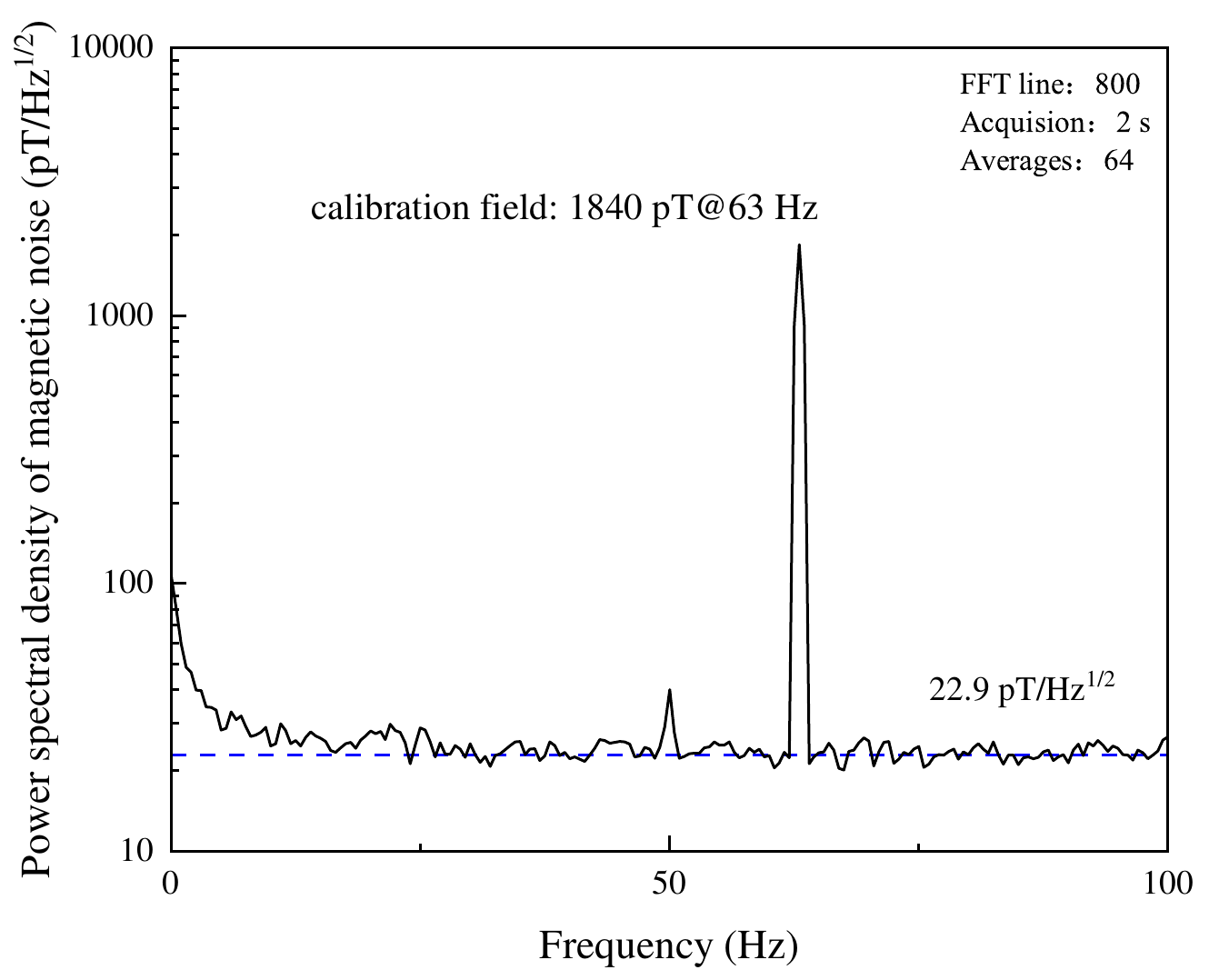} 
\caption{Power spectral density of magnetic noise of the magnetometer in closed-loop locked mode. The experimental conditions were identical to the open-loop test, with a calibration magnetic field of 63~Hz and 1.84~nT applied. Under the present test conditions, the measured noise floor is further reduced in the closed-loop state. The measured closed-loop sensitivity is 22.9~pT/Hz$^{1/2}$.}
\label{fig8}
\end{figure}

Following the verification of the system's superior dynamic locking capability, we further evaluated the sensitivity performance of the magnetometer under closed-loop conditions. A weak sinusoidal calibration magnetic field with a frequency of 63~Hz and an amplitude of 1.84~nT was applied. The measured Power Spectral Density (PSD) is illustrated in Fig.~\ref{fig8}. Spectral analysis results indicate that, under the present test conditions, the closed-loop locked state reduces the measured noise floor and leads to an enhancement of the signal-to-noise ratio (SNR) at the calibration frequency. Based on the noise floor calculation, the closed-loop sensitivity of the system reaches 22.9~pT/Hz$^{1/2}$.

\begin{figure}[htbp]
\centering
\includegraphics[width=0.55\linewidth]{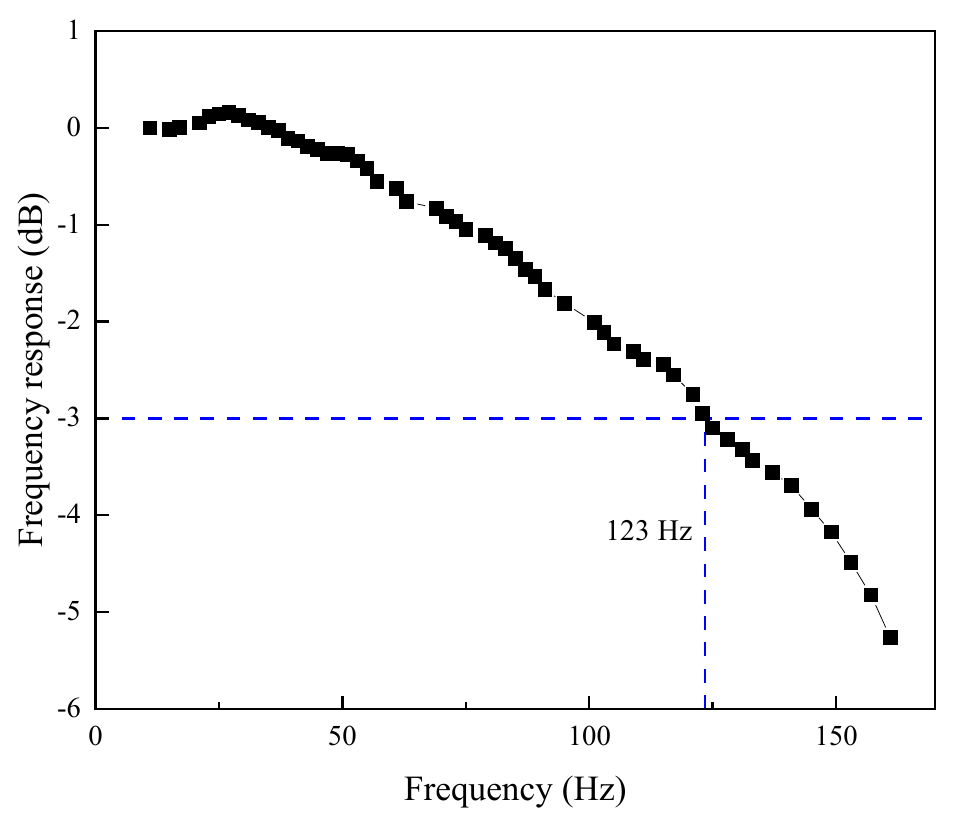} 
\caption{Closed-loop bandwidth test results of the magnetometer. A sinusoidal sweeping magnetic field with an amplitude of 1.8~nT was applied. The black curve represents the measured response amplitude (in dB) at different frequencies. The blue dashed line indicates the $-3$~dB attenuation point, determining the system's $-3$~dB bandwidth to be approximately 123~Hz.}
\label{fig9}
\end{figure}

Finally, the closed-loop bandwidth of the system was characterized using the frequency sweep method. A weak alternating calibration magnetic field with an amplitude of 1.8~nT was applied, and the variation of the FFT response peak was recorded. As shown in Fig.~\ref{fig9}, the response amplitude exhibits a gradual attenuation with increasing frequency. Defining the bandwidth as the frequency where the response amplitude decreases by 3~dB, the measured $-3$~dB bandwidth of the magnetometer is approximately 123~Hz. This measured bandwidth indicates that the system can track low-frequency magnetic-field variations within the tested frequency range.

\FloatBarrier

\subsection{Implementation of Vector Measurement}

\begin{figure}[htbp]
\centering
\includegraphics[width=0.9\linewidth]{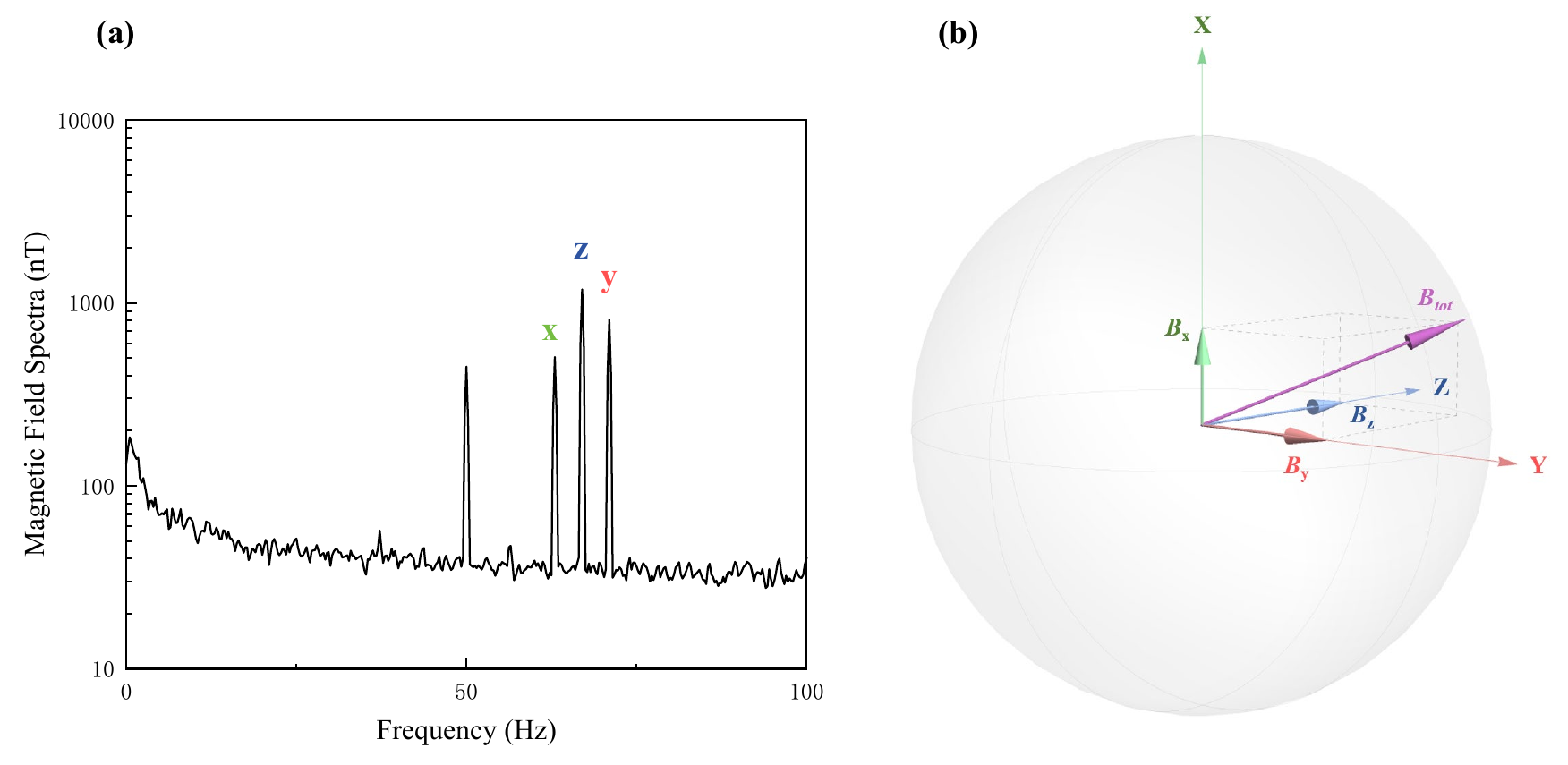} 
\caption{Experimental results in vector measurement mode. \textbf{(a)} Signal spectrum response. Sinusoidal modulation magnetic fields with an amplitude of 1~nT were applied along the $x$, $y$, and $z$ orthogonal axes at frequencies of 63~Hz, 71~Hz, and 67~Hz, respectively. The spectrum clearly displays three discrete signal peaks corresponding to the modulation frequencies. \textbf{(b)} 3D visualization of the reconstructed magnetic field vector. The static magnetic field components ($B_x, B_y, B_z$) are resolved from the amplitudes of the spectral peaks in \textbf{(a)}, and the resultant total magnetic field vector ($B_{\text{tot}}$) is synthesized in the Cartesian coordinate system, demonstrating the system's vector detection capability.}
\label{fig10}
\end{figure}

Following the achievement of high-sensitivity scalar magnetic field measurement and closed-loop locking, the vector measurement capability of the magnetometer was experimentally validated at the optimal operating point, based on the principle of tri-axial modulation. To realize the vectorization of the Mz magnetometer, weak low-frequency alternating magnetic fields were individually applied along the three orthogonal axes ($x, y, z$) to serve as modulation signals. To effectively avoid the influence of 50~Hz power line interference and its harmonics, while preventing crosstalk between axial signals, three non-harmonically related frequencies were selected for the experiment: 63~Hz for the $x$-axis, 71~Hz for the $y$-axis, and 67~Hz for the $z$-axis. To ensure that the modulation signals remained within the perturbation regime and to avoid inducing nonlinear interference in the detection of the main scalar magnetic field, the modulation amplitudes along all three directions were uniformly set to 1~nT. Given that the experiment was conducted within a magnetically shielded environment, to construct a non-zero magnetic field vector for measurement, specific DC bias magnetic fields were simultaneously applied via the tri-axial coils to simulate ambient vector components. The experimental results are depicted in Fig.~\ref{fig10}, which presents the FFT spectrum of the system output signal following the application of the tri-axial modulation fields. Spectral Analysis: Within the analysis bandwidth of 0--100~Hz, the spectral background clearly exhibits three discrete signal peaks. The center frequencies of these peaks correspond precisely to the preset modulation frequencies (63~Hz, 71~Hz, and 67~Hz). By extracting the peak amplitudes at these frequency points from the FFT spectrum and combining them with the known modulation depth, the magnetic field components currently applied along the $x, y$, and $z$ axes can be successfully retrieved. These experimental results fully demonstrate that, through coil modulation and the frequency-domain extraction algorithm, this study successfully realizes the functional extension from scalar measurement to vector measurement.

It should be noted that while the vectorization scheme based on auxiliary magnetic field modulation successfully enables the detection of magnetic field direction, the vector sensitivity inevitably suffers a degradation compared to the scalar sensitivity. According to the theoretical model of vectorization \cite{ref19}, the noise level in vector measurement is amplified by a factor of $B_0/\beta$, where $B_0$ denotes the background static magnetic field strength and $\beta$ represents the amplitude of the modulation magnetic field. In this experiment, with a background field of $B_0 = 1216$~nT and a modulation amplitude of $\beta = 1$~nT, the corresponding noise amplification factor is approximately 1216. Given the system's baseline scalar sensitivity of 22.9~pT/Hz$^{1/2}$, the vector measurement sensitivity at this operating point is estimated to be approximately 27.8~nT/Hz$^{1/2}$. This result indicates that, while maintaining the compact single-beam architecture, further improvement in vector detection precision relies on the further suppression of the scalar noise floor and the rational optimization of the modulation depth $\beta$, provided that the perturbation approximation condition is satisfied.

Although this sensitivity is slightly inferior to that of high-performance handheld tri-axial fluxgate magnetometers (e.g., MEDA FVM400 \cite{ref22}, resolution $\sim$1~nT), it is of a comparable order of magnitude to standard handheld fluxgate magnetometers (e.g., TUNKIA TM4300B \cite{ref23}, resolution $\sim$10~nT), demonstrating the capability for effective directional identification in geomagnetic environments. More importantly, unlike traditional fluxgate magnetometers that only provide nT-level detection capabilities, the proposed system integrates pT-level high-sensitivity scalar detection with nT-level vector detection functions, offering unique application advantages over standalone fluxgate magnetometers.

\section{Conclusion}
This paper presents the development of an $^{87}$Rb Mz-type atomic magnetometer based on a paraffin-coated anti-relaxation vapor cell. By employing multi-parameter joint optimization and closed-loop locking techniques, the system achieved scalar magnetic-field detection of $22.9~\mathrm{pT}/\mathrm{Hz}^{1/2}$ under near-room-temperature operation. Step magnetic-field response tests further indicate stable closed-loop tracking under the tested perturbation. By applying weak low-frequency modulation magnetic fields along three orthogonal axes and utilizing frequency-domain demodulation algorithms, we demonstrated vector magnetic measurement in a laboratory magnetic shielding environment. Owing to the intrinsic noise amplification associated with perturbative modulation, the current vector sensitivity remains at the nT level. Future work will focus on further optimizing the modulation and demodulation algorithms and improving three-axis coil calibration to enhance vector accuracy and sensitivity. Overall, this system provides an experimental basis for future development toward applications such as geomagnetic navigation and magnetic anomaly detection.

\begin{backmatter}

\bmsection{Funding}
This work was supported by the Basic Research Program of Shanxi Province (Grant No. 202403021211013), the National Natural Science Foundation of China (Grant No. 12474483), and the Teaching Instrument Development Project of School of Physics and Electronic Engineering, Shanxi University.

\bmsection{Disclosures}
The authors declare no conflicts of interest.

\bmsection{Data availability}
Data underlying the results presented in this paper are available from the corresponding author upon reasonable request.

\end{backmatter}






\begin{thebibliography}{99}

\providecommand{\enquote}[1]{``#1''}

\bibitem{ref1}
M.~S.~Mrozowski, A.~S.~Bell, P.~F.~Griffin, D.~Hunter, D.~Burt, J.~P.~McGilligan, E.~Riis, C.~D.~Beggan, and S.~J.~Ingleby, \enquote{Distributed network of optically pumped magnetometers for space weather monitoring,} {\protect\JournalTitle{Sci. Rep.}} \textbf{14}, 28229 (2024).

\bibitem{ref2}
T.~S.~Horbury, H.~O'Brien, I.~Carrasco~Blazquez, M.~Bendyk, P.~Brown, R.~Hudson, V.~Evans, T.~M.~Oddy, C.~M.~Carr, T.~J.~Beek, E.~Cupido, S.~Bhattacharya, J.~A.~Dominguez, L.~Matthews, V.~R.~Myklebust, B.~Whiteside, S.~D.~Bale, W.~Baumjohann, D.~Burgess, and A.~P.~Walsh, \enquote{The solar orbiter magnetometer,} {\protect\JournalTitle{Astron. Astrophys.}} \textbf{642}, A9 (2020).

\bibitem{ref3}
Y.~Lu, T.~Zhao, W.~Zhu, L.~Liu, X.~Zhuang, G.~Fang, and X.~Zhang, \enquote{Recent progress of atomic magnetometers for geomagnetic applications,} {\protect\JournalTitle{Sensors}} \textbf{23}, 5318 (2023).

\bibitem{ref4}
X.~Zhang, C.~Q.~Chen, M.~K.~Zhang, C.~Y.~Ma, Y.~Zhang, H.~Wang, Q.~Q.~Guo, T.~Hu, Z.~B.~Liu, Y.~Chang, K.~J.~Hu, and X.~D.~Yang, \enquote{Detection and analysis of MEG signals in occipital region with double-channel OPM sensors,} {\protect\JournalTitle{J. Neurosci. Methods}} \textbf{346}, 108948 (2020).

\bibitem{ref5}
T.~Wu, W.~Xiao, X.~Peng, T.~Wu, and H.~Guo, \enquote{Compact high-bandwidth single-beam optically-pumped magnetometer for biomagnetic measurement,} {\protect\JournalTitle{Opt. Express}} \textbf{16}, 235 (2024).

\bibitem{ref6}
D.~Budker and M.~Romalis, \enquote{Optical magnetometry,} {\protect\JournalTitle{Nature Phys.}} \textbf{3}, 227 (2007).

\bibitem{ref7}
A.~Fabricant, I.~Novikova, and G.~Bison, \enquote{How to build a magnetometer with thermal atomic vapor: a tutorial,} {\protect\JournalTitle{New J. Phys.}} \textbf{25}, 025001 (2023).

\bibitem{ref8}
S.~Groeger, G.~Bison, J.~L.~Schenker, R.~Wynands, and A.~Weis, \enquote{A high-sensitivity laser-pumped Mx magnetometer,} {\protect\JournalTitle{Eur. Phys. J. D}} \textbf{38}, 239 (2006).

\bibitem{ref9}
Y.~Gu, R.~Shi, and Y.~Wang, \enquote{Study on sensitivity-related parameters of distributed feedback laser pumped rubidium magnetometer,} {\protect\JournalTitle{Acta Physica Sinica}} \textbf{63}, 110701 (2014). (in Chinese)

\bibitem{ref10}
L.~Zhang, L.~Bai, Y.~Yang, Y.~Yang, Y.~Wang, X.~Wen, J.~He, and J.~Wang, \enquote{Improving the sensitivity of optically pumped rubidium atomic magnetometer by using repumping light ,} {\protect\JournalTitle{Acta Physica Sinica}} \textbf{70}, 230702 (2021).(in Chinese)

\bibitem{ref11}
K.~Xu, X.~Ren, Y.~Xiang, M.~Zhang, X.~Zhao, K.~Ma, Y.~Tian, D.~Wu, Z.~Zeng, and G.~Wang, \enquote{Multi-parameter optimization of rubidium laser optically pumped magnetometers with geomagnetic field intensity,} {\protect\JournalTitle{Sensors}} \textbf{23}, 8919 (2023).

\bibitem{ref12}
C.~Zhan, Z.~Ma, J.~Wu, M.~Li, C.~Han, B.~Lu, and C.~Lee, \enquote{Magnetic field stabilization system designed for the cold-atom coherent population-trapping clock,} {\protect\JournalTitle{Chin. Opt. Lett.}} \textbf{22}, 080202 (2024).

\bibitem{ref13}
R.~Zhu, Y.~Zhang, P.~Du, Y.~Xuan, F.~Yang, Y.~Zhou, K.~Zhang, X.~Li, S.~Sun, T.~Cheng, J.~Li, S.~Zheng, W.~Quan, and J.~Li, \enquote{A compact optically pumped potassium atomic magnetometer with high sensitivity under geomagnetic field intensity,} {\protect\JournalTitle{Measurement}} \textbf{250}, 117099 (2025).

\bibitem{ref14}
H.~Zhai, W.~Li, and G.~Jin, \enquote{Improving the sensitivity of a dark-resonance atomic magnetometer,} {\protect\JournalTitle{Sensors}} \textbf{25}, 1229 (2025).



\bibitem{ref15}
C.~Cohen-Tannoudji, J.~Dupont-Roc, S.~Haroche, and F.~Lalo\"e, \enquote{Diverses r\'esonances de croisement de niveaux sur des atomes pomp\'es optiquement en champ nul. I. Th\'eorie,} {\protect\JournalTitle{Rev. Phys. Appl. (Paris)}} \textbf{5}, 95 (1970).

\bibitem{ref16}
W.~Xiao, Y.~Wu, X.~Zhang, Y.~Feng, C.~Sun, T.~Wu, J.~Chen, X.~Peng, and H.~Guo, \enquote{Single-beam three-axis optically pumped magnetometers with sub-100 femtotesla sensitivity,} {\protect\JournalTitle{Appl. Phys. Express}} \textbf{14}, 066002 (2021).

\bibitem{ref17}
R.~Dawson, M.~S.~Mrozowski, D.~Hunter, C.~O'Dwyer, E.~Riis, P.~F.~Griffin, and S.~Ingleby, \enquote{A triaxial vectorization technique for a single-beam zero-field atomic magnetometer to suppress cross-axis projection error,} arXiv Preprint arXiv:2408.12994 (2024).

\bibitem{ref18}
Y.~Zou, L.~Jiang, H.~Bai, J.~Liu, C.~Fang, J.~Zhu, Q.~Shao, J.~Xu, X.~Zhou, and W.~Quan,
\enquote{Vector magnetometry employing a rotating RF field in a single-beam optically pumped magnetometer,} {\protect\JournalTitle{Sensors Actuators A}} \textbf{379}, 115009 (2024).

\bibitem{ref19}
C.~T.~Chism, R.~A.~Marshall, A.~Aguilar-Nadalini, S.~Knappe, D.~Malaspina, K.~J.~Hughes, M.~Ellmeier, S.~Wankmueller, C.~Cunningham, T.~Maydew, and O.~Alem, \enquote{Design of a vectorized rubidium scalar magnetometer for SmallSat applications,} {\protect\JournalTitle{Adv. Space Res.}} \textbf{76}, 4635 (2025).

\bibitem{ref20}
T.~Wang, W.~Lee, M.~Limes, T.~Kornack, E.~Foley, and M.~Romalis, \enquote{Pulsed vector atomic magnetometer using an alternating fast-rotating field,} {\protect\JournalTitle{Nature Commun.}} \textbf{16}, 1374 (2025).

\bibitem{ref21}
Q.~Gan, J.~Shang, Y.~Ji, and L.~Wu,
\enquote{Simultaneous excitation of $^{85}$Rb and $^{87}$Rb isotopes inside a microfabricated vapor cell with double-RF fields for a chip-scale Mz magnetometer,} {\protect\JournalTitle{Rev. Sci. Instrum.}} \textbf{88}, 115009 (2017).

\bibitem{ref22}
MEDA, Inc., \enquote{FVM400 Vector Magnetometer Instruction Manual,} Dulles, VA, USA (2018).

\bibitem{ref23}
TUNKIA Co., Ltd., \enquote{TM4300B Hand-held Three-axis Fluxgate Magnetometer User Manual,} Changsha, China.

\end{thebibliography}
\end{document}